\documentclass[prb,twocolumn]{revtex4}

\usepackage{amsmath}
\usepackage{graphicx}
\usepackage{times}

\begin{document}

\title{Ferromagnetism of Weakly-Interacting Electrons in Disordered Systems}

\author{Xiao Yang$^1$ and Chetan Nayak$^2$}

\affiliation{$^1$ Physics Department, University of Virginia, Charlottesville, VA 22904-4714\\
$^2$Department of Physics and Astronomy,
University of California at Los Angeles, Los Angeles,
CA 90095-1547}

\date{\today}

\begin{abstract}
It was realized two decades ago that the two-dimensional diffusive
Fermi liquid phase is unstable against arbitrarily weak
electron-electron interactions.
Recently, using the nonlinear sigma model developed by
Finkelstein, several authors have shown
that the instablity leads to a ferromagnetic state. In this
paper, we consider diffusing electrons interacting through
a ferromagnetic exchange interaction.
Using the Hartree-Fock
approximation to directly calculate the electron self
energy, we find that the total energy is
minimized by a finite ferromagnetic moment
for arbitrarily weak interactions in two dimensions
and for interaction strengths exceeding a critical
value proportional to the conductivity in three dimensions.
We discuss the relation between our results and previous ones.
\end{abstract}

\maketitle

\section{Introduction}
Since the discovery of Anderson localization, it has been well
known that disorder can profoundly change the electronic properties
of solids. For non-interacting electronic systems, the celebrated scaling
theory of Abrahams et.al \cite{Abrahams} shows that
electrons are always localized in one and two dimensions, while in
three dimensions, there exists a metal-insulator transition at
a critical value of the disorder strength. On the other hand, when interactions
between electrons are taken into account, the results are much
less clear. In a disordered system, at length scales beyond the
mean free path, electrons move diffusively. Since diffusion is
slow, electron-electron interactions are consequently enhanced.
First, electrons' ability to screen the long-ranged Coulomb interaction
is hampered. Second, electrons stay in each
other's region longer and hence interact with each other more
strongly \cite{Altshuler}. As shown by Finkelstein \cite{Finkelstein} and
Castellani {\it et al.} \cite{Castellani}, the two-dimensional
diffusive Fermi liquid phase is, as a result, unstable. However, it is
not clear what is the stable fixed point controlling the low-energy behavior
of weakly-interacting electrons in the presence of disorder.

Using the nonlinear sigma model developed by
Finkelstein \cite{Finkelstein}, several
authors \cite{Belitz,Chamon,Nayak} have shown that the enhancement
of electron-electron interactions due to disorder leads to the formation of
ferromagnetic moment. The non-linear sigma model takes as its starting point
a repulsive density-density interaction which is customarily split into small-angle
and large angle scattering with coupling constants $\Gamma_1$ and $\Gamma_2$
\cite{Finkelstein}. The latter coupling drives ferromagnetism.
In this paper, we consider itinerant electrons interacting through a
ferromagnetic exchange $H_{\rm int}=-\int J({\bf r}-{\bf r'}){\bf S}(r)\cdot {\bf S}(r')$.
As in the case of a density-densty interaction, this is an interaction
which, in the clean, disorder-free case, does not lead to ferromagnetism
for small $J$; if it occurs at all, it can only be at large $J$.
This is because a kinetic energy penalty must be paid when
the up- and down-spin Fermi surfaces are split in order
generate a net ferromagnetic moment. The kinetic energy penalty
for splitting the Fermi surfaces by $\delta\mu$ is $\sim \nu\, \delta\mu^2$,
where $\nu$ is the density of states at the Fermi surface. On the other
hand, the energy gain for non-interacting electrons is $\sim J {\nu^2}\, \delta\mu^2$.
Hence, ferromagnetism can only occur for $J>\frac{1}{\nu}$, the Stoner criterion.
However, as we shall see in the next section, in
a two-dimensional disordered system, the interaction energy gain
varies instead as $\sim J {\nu^2}\, {\delta\mu^2}\,\ln(\tau\, {\delta\mu})$;
consequently, it is always favorable to develop a ferromagnetic moment.
(This is similar to what occurs in a quantum dot\cite{Andreev98}.)
We show this by directly calculating the disorder-averaged
electron self-energy, using the trick of ref. \onlinecite{Abrahams81}, in which
it is related to the diffusive pole in the density-density correlation
function. This method is physically transparent since one can directly see how the
seemingly innocent diffusive pole leads to a logarithmic
correction to the self energy. We also avoid the need for the replica
trick used in the sigma model treatment. For a three dimensional system, our
calculation shows that ferromagnetic order develops when
$J$ exceeds a critical value which is proportional to the
conductivity of the system. We compare these results with previous ones.

\section{The Model}
We consider the following Hamiltonian for a weakly-interacting
electronic system in the presence of a random distribution of
impurities:
\begin{equation}
H = H_{0} + H_{\rm int}
\end{equation}
where
\begin{equation}
H_{0} = \sum_{m,\alpha}E_{m}\,\hat{c}^{\dagger}_{m\alpha}\hat{c}^{}_{m\alpha}
\end{equation}
is the noninteracting Hamiltonian, which includes the kinetic
energy of electrons and the random potential. $E_{m}$, ${\varphi_m}(r)$ are
respectively the $m^{\rm th}$ eigenvalue and eigenfunction of $H_0$,
which we assume to be spin-independent. We will call $E_m$ the `kinetic' energy
of the electron to distinguish it from the interaction energy associated
with $H_{\rm int}$, even though $E_m$ includes the effects of the random potential.
$c_{m \alpha}$ is the annihilation operator for an electron of
spin $\alpha=\uparrow,\downarrow$ in the eigenstate ${\varphi_m}(r)$.
The annihilation operator $\psi_{\alpha}(r)$ for an electron of spin $\alpha$ at ${\bf r}$ is:
\begin{equation}
\psi_{\alpha}(r) = \sum_{m}\varphi_{m}(r)\,\hat{c}^{}_{m\alpha}
\end{equation}
As mentioned in the introduction, the
interaction Hamiltonian $H_{\rm int}$ considered here is:
\begin{multline}
H_{\rm int} = -\int {d^d}{\bf r}\,{d^d}{\bf r'}\, J({\bf r}-{\bf r'})
{\bf S}({\bf r}) \cdot {\bf S}({\bf r'})\\
=-\int {d^d}{\bf r}\,{d^d}{\bf r'}\, J({\bf r}-{\bf r'}) \bigl( S^{z}({\bf r})S^{z}({\bf r'})\\
+ 2\left(S^{+}({\bf r})S^{-}({\bf r'})+
S^{-}({\bf r})S^{+}({\bf r'})\right)\bigr)
\end{multline}
Here $J({\bf r}-{\bf r'})>0$ is the exchange interaction which
is assumed to be weak, ferromagnetic, and short-ranged.
${\bf S(r)}=\hat{\psi}^{\dagger}_{\alpha}(r)\vec{\sigma}_{\alpha\beta}\hat{\psi}_{\beta}(r)$
is the spin density operator, and $\vec{\sigma}$ are the Pauli matrices.

\section{Hartree and Fock Energies}

As a warmup, let us first consider a clean system, for which the
${\varphi_m}(r)$ are plane waves. We look for ferromagnetism in this
model by assuming taking a trial wavefunction $\left|\Psi_0\right\rangle_{\rm clean}$
with a ferromagnetic moment:
\begin{equation}
\left|{\Psi_0}\right\rangle_{\rm clean} =
\prod_{k<k_{F\uparrow}}\prod_{k'<k_{F\downarrow}}
{c^\dagger_{k\uparrow}}{c^\dagger_{k'\downarrow}}
\left|0\right\rangle
\end{equation}
and minimizing $\left\langle{\Psi_0}\right| H \left|{\Psi_0}\right\rangle$
with respect to the magnetization $M= \nu\,\delta\!\mu$,
where $\nu$ is the single-electron density of states and
$2\,\delta\!\mu=\bigl(k^2_{F\uparrow}-k^2_{F\downarrow}\bigr)/2m$
is the difference in the kinetic energies of up- and down-spin
electrons at the Fermi energy.

In evaluating the expectation value of the interaction energy,
we find two terms:
\begin{multline}
\left\langle{\Psi_0}\right| H_{\rm int} \left|{\Psi_0}\right\rangle =
-J\!\left(0\right) {M^2}\,+\\
{\sum_{\bf k,k'}}J\!\left({\bf k}-{\bf k'}\right)
\left(\frac{3}{8}{N_k}{N_{k'}}-\frac{1}{2}{S^z_k}{S^z_{k'}}\right)
\end{multline}
where ${N_k} , \, {S^z_k} \: = \: \theta(k_{F\uparrow}-k)\pm\theta(k_{F\downarrow}-k)$.
They are essentially the Hartree and Fock contributions to the energy.

If we take for simplicity $J\!\left({\bf q}\right)=J$, then
\begin{equation}
\label{eqn:pot-energy-clean}
\left\langle H_{\rm int}  \right\rangle = -\frac{3}{2}J{M^2}+\frac{3}{8}J{N^2}
\end{equation}
Ferromagnetism will only occur if this interaction energy
is larger than the kinetic energy cost
\begin{equation}
\left\langle H_{\rm kin}  \right\rangle_{M} - \left\langle H_{\rm kin}  \right\rangle_{0}
= \nu \, {\delta\!\mu^2} = \frac{1}{\nu}\,{M^2}
\end{equation}
i.e. only for $3J/2>\nu$. (The magnetization is stabilized at a finite
value by the energy dependence of the density-of states, which we have
neglected here.)

Consider, instead, the dirty case. We take
\begin{equation}
\left|{\Psi_0}\right\rangle_{\rm dirty} =
\prod_{n<{n_0}+M}\: \prod_{n'<{n_0}-M}
{c^\dagger_{n\uparrow}}{c^\dagger_{n'\downarrow}}
\left|0\right\rangle
\end{equation}
where $E_{n_0}={E_F}$. Again, the magnetization $M$ is proportional
to the difference in the `kinetic' energies of up- and down-spin electrons at
the Fermi level: $M=\nu \,\delta\mu$, where $\nu$ is the single-electron
density of states. Now, we find that the disorder-averaged variational
interaction energy is:
\begin{multline}
\label{eqn:dirty-var-energy}
\overline{\left\langle{\Psi_0}\right| H_{\rm int} \left|{\Psi_0}\right\rangle} =
-\int {d^d}{\bf r}\,{d^d}{\bf r'}\, J\!\left({\bf r}-{\bf r'}\right) 
\overline{\left\langle{S^z}(r)\right\rangle \left\langle{S^z}(r')\right\rangle}\\
+\, \int dE\,dE'\int {d^d}{\bf r}\,{d^d}{\bf r'}\, J\!\left({\bf r}-{\bf r'}\right)
F\!\left(E,E',{\bf r},{\bf r'}\right)\,{N_E}N_{E'}
\end{multline}
where
\begin{equation}
\left\langle{S^z}(r)\right\rangle =
{\sum_{m}} \varphi^{*}_{m}(r)\varphi_{m}(r) \left\langle N_{m\uparrow}-N_{m\downarrow}\right\rangle
\end{equation}
\begin{equation}
\left\langle N_{m\uparrow,\downarrow}\right\rangle = \left\langle {c^\dagger_{m\uparrow,\downarrow}}{c^{}_{m\uparrow,\downarrow}}\right\rangle
=\theta({n_0}\pm M-m)
\end{equation}
$N_E$ is the number of electrons in the state at
energy $E$, namely ${N_E}=2$ for $E<-\delta\!\mu$,
${N_E}=1$ for $-\delta\!\mu<E<\delta\!\mu$, and 
${N_E}=0$ for $E>\delta\!\mu$.
The function $F(E,E';r,r')$ is defined as
\begin{multline}
F(E,E';r,r')=\sum_{m,n}\delta(E-E_m)\delta(E'-E_n)\, \times\\
\overline{\varphi^{*}_{m}(r)\varphi_{m}(r')\varphi^{*}_{n}(r')\varphi_{n}(r)}\:
\end{multline}
The upper bar indicates average over disorder.
The first term in (\ref{eqn:dirty-var-energy}) is similar in behavior
to the clean case, and will be neglected in the following.
The second term is more interesting because
the function $F$ is closely related to the density-density
correlation function. 
The crucial difference with the clean case is that
the latter has a diffusive form, as we now describe.

We assume that the density propagates diffusively, so that
\begin{eqnarray}
\Pi({\bf q},\omega)&=&i\int^{\infty}_{0}dt\,{d^d}{\bf x}\,\overline{\left\langle[\rho({\bf x},t),\rho(0,0)]\right\rangle}e^{-i{\bf q}\cdot{\bf x}+i\omega\!t}\cr
&=&\nu\frac{Dq^2}{-i\omega +Dq^2}
\end{eqnarray}
where, again, $\nu$ is the density of states at the Fermi surface, and $D$ is the diffusion
constant. Then
\begin{multline}
\Pi^{*}({\bf q},\omega) = -i\int^{\infty}_{0}dt\,{d^d}{\bf x}\,\overline{\left\langle[\rho(0,0),\rho({\bf x},t)]\right\rangle}e^{i{\bf q}\cdot{\bf x}-i\omega t}\\
=-i\int_{-\infty}^{0}dt\,{d^d}{\bf x}\,\overline{\left\langle[\rho({\bf x},t),\rho(0,0)]\right\rangle}
e^{-i{\bf q}\cdot{\bf x}+i\omega t}
\end{multline}
On the second line, we have used the fact that after averaging
over disorder, the density-density correlation function is
translationally invariant. This assumption is valid in the high
conductivity regime of interest here.
From this, we can obtain the Fourier transform of the density-density correlation function,
\begin{equation}
A({\bf q},\omega)=\int^{\infty}_{-\infty}dt\, {d^d}{\bf x}\,
e^{-i{\bf q}\cdot{\bf x}+i\omega t}\overline{\left\langle[\rho({\bf x},t),\rho(0,0)]\right\rangle}
\end{equation}
by observing that
\begin{eqnarray}
A({\bf q},\omega)&=&\frac{1}{i}(\Pi({\bf q},\omega)-\Pi^{*}({\bf q},\omega))\\
\nonumber &=&\frac{1}{2}\,{\rm Im}(\Pi({\bf q},\omega))\\ \nonumber
&=&\nu\frac{Dq^2\omega}{2(\omega^2+(Dq^2)^2)}
\end{eqnarray}

We can relate this to our function $F$ by noting that
\begin{eqnarray}
\left\langle[\rho(rt),\rho(r',0)]\right\rangle=
\sum_{m,n}\varphi^{*}_{m}(r)\varphi_{m}(r')\varphi^{*}_{n}(r')\varphi_{n}(r)\cr
\times \left\langle N_{m}-N_{n} \right\rangle
\end{eqnarray}
So we get
\begin{eqnarray}
A({\bf q},\omega) &=& \int_{\mu}^{\infty}dE\int_{-\infty}^{\mu}dE' \: F(E,E';r,r')\cr
& & {\hskip 2 cm} \times\: e^{iq(r-r')}\delta(E-E'-\omega)\cr
&=& \omega F({\bf q},\omega)
\end{eqnarray}
Where, on the last line, $F({\bf q},\omega)$ is the Fourier transform of
$F(E,E';r,r')$, and $\omega$ is the energy difference $E-E'$. $\mu$ is
the chemical potential. So we get
\begin{equation} \label{eq:F}
F({\bf q},\omega)=\nu\frac{Dq^2}{2(\omega^2+(Dq^2)^2)}
\end{equation}
and the second term in (\ref{eqn:dirty-var-energy}) becomes
\begin{multline}
\label{eq:selfenergy}
\overline{\left\langle{\Psi_0}\right| H_{\rm int} \left|{\Psi_0}\right\rangle} = \ldots +\\
\nu\int dE \int^{\mu}_{-\infty}dE' \int
\frac{d^dq}{(2\pi)^d} \frac{Dq^2}{(E-E')^2+(Dq^2)^2}J({\bf q})
\end{multline}
The $\ldots$ refers to the first term in (\ref{eqn:dirty-var-energy}),
which is smaller because it does not have any infrared singularities in
its integral.
In the preceding derivation, we 
performed the disorder average of $F$, by comparing it to the
density-density correlation function, thereby avoiding
the replica trick normally used in the field theoretical treatment.

\section{Ferromagnetic Ground State}

We now minimize $\langle H\rangle$ with respect to the
magnetization $M$. We define the effective interaction
$V(E-E')$ between electrons in
states at energies $E$, $E'$ by
\begin{equation}
\overline{\left\langle{\Psi_0}\right| H_{\rm int} \left|{\Psi_0}\right\rangle} =
\!\int_{E>E'} dE \,dE' \, V(E-E')\,{N_E}N_{E'}
\end{equation}
Evaluating the momentum integral in (\ref{eq:selfenergy}), we
obtain
\begin{equation}
\label{eq:selfenergy2}
V(E-E')=\frac{{\nu} J}{8\pi D}\: \ln\!\left[\frac{(E-E')^2+\Lambda^2}{(E-E')^2}\right]
\end{equation}
Here, we have ignored the momentum dependence of $J({\bf q})$,
and as usual, $\Lambda$ is the ultraviolet cutoff which can be taken to
be $1/\tau$, where $\tau$ is the elastic scattering time.

Thus, the interaction energy is:
\begin{multline}
\overline{{\left\langle H_{\rm int}  \right\rangle_M}} = \int^{-\delta\!\mu}_{-\Lambda}dE
\int^{E}_{-\Lambda} dE' \:V(E-E')\cdot2\cdot 2\\
+  \int^{\delta\!\mu}_{-\delta\!\mu}dE
\int^{-\delta\!\mu}_{-\Lambda} dE' \:V(E-E')\cdot 2\cdot 1\\
+  \int^{\delta\!\mu}_{-\delta\!\mu}dE
\int^{E}_{-\delta\!\mu} dE' \:V(E-E')\cdot 1\cdot 1
\end{multline}
Subtracting the interaction energy in the paramagnetic state, $M=0$,
\begin{equation}
\overline{{\left\langle H_{\rm int}  \right\rangle_{0}}} = \int^{0}_{-\Lambda}dE
\int^{E}_{-\Lambda} dE' \:V(E-E')\cdot2\cdot 2
\end{equation}
we have
\begin{multline}
\overline{{\left\langle H_{\rm int}  \right\rangle_M}}-
\overline{{\left\langle H_{\rm int}  \right\rangle_{0}}}
 = -\int^{\delta\!\mu}_{-\delta\!\mu}dE
\int^{E}_{-\delta\!\mu} dE' \:V(E-E')\\
= -\frac{{\nu}J}{8\pi D}\int^{\delta\!\mu}_{-\delta\!\mu}dE
\int^{E}_{-\delta\!\mu} dE'\:  \ln\!\left[\frac{(E-E')^2+\Lambda^2}{(E-E')^2}\right]\\
= -\frac{{\nu}J}{8\pi D}
\int^{\delta\!\mu}_{-\delta\!\mu}dE\,(E+\delta\!\mu)
\left(1+\ln\!\left[\frac{(E+\delta\!\mu)^2+\Lambda^2}{(E+\delta\!\mu)^2}\right]\right)\\
= -\frac{{\nu}J}{8\pi D}\left( 6\, {\delta\!\mu^2} +
4\, {\delta\!\mu^2} \ln\!\left(\frac{\Lambda}{2\delta\!\mu}\right)\right)
\end{multline}
Including the `kinetic energy', we have
\begin{multline}
\overline{{\left\langle H \right\rangle_M}} -
\overline{{\left\langle H  \right\rangle_{0}}}
 = \left(\nu -\frac{3{\nu}J}{4\pi D}\right)\, {\delta\!\mu^2} 
-\frac{{\nu}J}{2\pi D}\, {\delta\!\mu^2} \ln\!\left(\frac{\Lambda}{2\delta\!\mu}\right)
\end{multline}

Even at small $J$, ferromagnetic order occurs because the
total energy is minimal at a finite value of $\delta\!\mu$ and, therefore,
of $M$:
\begin{equation}
M = \frac{\nu \Lambda}{2} \, e^{-{2\pi D}/{J}}
\end{equation}
So we have shown that the diffusive fermi liquid is always
unstable against the ferromagnetic order in two dimensions.

In three dimensions, one can do the analogous calculations, leading to
\begin{equation}
\left\langle H \right\rangle - {E_0} = \left(\nu-{\nu^2}J-\frac{J\nu\Lambda^{1/2}}{2\pi^2D^{3/2}}\right)
{\delta\!\mu^2} + \frac{4J\nu}{15\pi D^{3/2}}{\delta\!\mu^{5/2}}
\end{equation}
In the first term, we have included the Hartree contribution because the
Fock part is non-divergent and can be viewed as a correction.
Hence, the critical coupling $J_c$ above which ferromagnetism
occurs is modified from the Stoner value:
\begin{equation}
{J_c} = \frac{1}{\nu}\left(1 - \frac{3\sqrt{3}}{\left({k_F}\ell\right)^2}\right)
\end{equation}
where $\ell={v_F}\tau$ is the elastic mean-free path and we have
used $D={v_F^2}\tau/3$ in three dimensions.

\section{discussion}

In this paper, we use a straightforward Fock approximation
to show that in two dimensions an arbitrarily weak ferromagnetic
interaction causes an instability of the diffusive Fermi liquid.
In three dimensions, this instability occurs only when the interaction
strength exceeds a critical value proportional to the
conductivity. We worked directly with the electronic Hamiltonian,
without recourse to Finkelstein's non-linear $\sigma$ model,
which was used at an intermediate stage of previous
calculations\cite{Belitz,Chamon,Nayak}. Our results are
consistent with these earlier ones.
The model and method of this paper has the
advantage of physical transparency: it shows the
direct connection between diffusive motion and the
ferromagnetic instability.

By using the method in ref. \onlinecite{Abrahams81},
we avoided using the replica trick, and saw the crucial
feature of diffusive motion: it implies a spatial correlation
between single-particle states which are nearby in energy.
As a result of this correlation, the effective interaction strength
diverges as the energy of two electrons approach each other.
This strongly favors ferromagnetism since electrons of the same
spin cannot occupy the same energy level. This strong dependence on
the energy separation of the electrons is completely absent in a
clean system.

We note that once ferromagnetic order develops, the system
goes into the universality class of electrons in a Zeeman field
\cite{Castellani}, which flows towards an insulating state in two dimensions.

\acknowledgements
We would like to thank C. Chamon and A. Kamenev for their comments
on an earlier version of this manuscript. C.N. was supported by
the National Science Foundation under Grant No. DMR-0411800. X.Y. was
supported by the Chemical Sciences, Geosciences and Biosciences Division,
Office of Basic Energy Sciences, Office of Sciences, U.S. Department of
Energy, and by NSF under Grant No. DMR-0412936.


\begin{thebibliography}{99}

\bibitem{Abrahams} E. Abrahams, P. W. Anderson, D. C.
Licciardello, and T. V. Ramakrishnan, Phys. Rev. Lett. {\bf 42},
673 (1979).


\bibitem{Altshuler} B. L. Altshuler and A. G. Aronov. Solid State Comm. {\bf
39}, 115 (1979); Sov. Phys. JETP {\bf 50}, 968 (1979).

\bibitem{Finkelstein} A. M. Finkelstein, Zh. Eksp. Teor. Fiz. {\bf
84}, 168 (1983), Sov. Phys. JETP {\bf 57}, 97 (1983).

\bibitem{Castellani} C. Castellani, C. Di Castro, P. A. Lee and M. Ma,
Phys. Rev. B {\bf 30} 527 (1984).

\bibitem{Belitz} D. Belitz and T. R. Kirkpatrick,
Phys. Rev. B {\bf 53}, 14364 (1996).

\bibitem{Chamon} C. Chamon and E. Mucciolo, Phys. Rev. Lett.
{\bf 85}, 5607 (2000).

\bibitem{Nayak} C. Nayak and X. Yang, Phys. Rev. B {\bf 68}, 104423 (2003).

\bibitem{Andreev98} A.~V. Andreev and A. Kamenev, \prl {\bf 81}, 3199 (1998).

\bibitem{Abrahams81} E. Abrahams, P.~W. Anderson, P.~A. Lee, and T.~V. Ramakrishnan,
Phys. Rev. B {\bf 24}, 6783 (1981).


\end{thebibliography}
\end{document}